\documentstyle[11pt,newpasp,twoside]{article}
\def\edcomment#1{\iffalse\marginpar{\raggedright\sl#1\/}\else\relax\fi}
\reversemarginpar

\begin{document}
\title{When Do Planets Form? A Search for Extra-solar Planets Around 
Metal-Poor Stars}
% \author{A. Sozzetti, D. W. Latham, G. Torres, \& R. P. Stefanik}
%\affil{Harvard-Smithsonian Center for Astrophysics, Cambridge, MA 02138}
%\affil{Harvard-Smithsonian Center for Astrophysics, 60 Garden Street, 
%Cambridge, MA 02138}
%\author{A. P. Boss}
%\affil{Carnegie Institution of Washington, Washington, DC 20015}
%\affil{Carnegie Institution of Washington, 5241 Broad Branch Road, 
%Washington, DC 20015}
%\author{B. W. Carney}
%\affil{University of North Carolina, Chapel Hill, NC 27599}
%\author{J. B. Laird}
%\affil{Bowling Green State University, Bowling Green, OH 43403}

\author{A. Sozzetti$^1$, D. W. Latham$^1$, G. Torres$^1$, R. P. Stefanik$^1$, 
A. P. Boss$^2$, B. W. Carney$^3$, \& J. B. Laird$^4$}
%\affil{Harvard-Smithsonian Center for Astrophysics, Cambridge, MA 02138}
\affil{$^1$Harvard-Smithsonian Center for Astrophysics, 60 Garden Street, 
Cambridge, MA 02138}
%\author{A. P. Boss}
%\affil{Carnegie Institution of Washington, Washington, DC 20015}
\affil{$^2$Dept. of Terrestrial Magnetism, 
Carnegie Institution of Washington, 5241 Broad Branch Road, 
Washington, DC 20015}
%\author{B. W. Carney}
\affil{$^3$Dept. of Physics and Astronomy, 
University of North Carolina, CB 3255 Phillips Hall, Chapel Hill, NC 27599}
%\author{J. B. Laird}
\affil{$^4$Dept. of Physics and Astronomy, Bowling Green State University, 
104C Overman Hall, Bowling Green, OH 43403}

\begin{abstract}
We present preliminary results from our spectroscopic search for 
planets within 1 AU of metal-poor field dwarfs using NASA time with HIRES on 
Keck I. The core accretion model of gas giant planet formation is sensitive to the 
metallicity of the raw material, while the disk instability model is not. 
By observing metal-poor stars in the field we eliminate the role of dynamical 
interactions in dense stellar environments, such as a globular cluster. 
The results of our survey should allow us to distinguish the relative roles of 
the two competing giant planet formation scenarios. 
\end{abstract}

\section{Introduction}

The two proposed models for the formation of gas giant planets make different, 
and testable, predictions. For example, core accretion requires several Myr 
to grow a solid core massive enough to accrete a gaseous envelope 
(Lissauer~1993). In contrast, disk instability leads to 
Jupiter-mass clumps (e.g., Boss~1997), which can survive and give rise to 
actual protoplanets (Mayer et al.~2002), within $\sim 10^3$ yr. 
The time-scale for core accretion to proceed depends strongly on the 
initial surface density of solids (Pollack et al. 1996), so that formation 
by this mechanism may be enhanced in metal-rich stars. 
Instead, disk instability is remarkably insensitive to
the primordial metallicity of the protoplanetary disk (Boss 2002). 
Several observational tests can be performed to probe giant planet 
formation models (Boss 2003). In order to unambiguously establish whether high 
primordial metallicity is a requirement, and thus determine which 
is the dominant formation mechanism for 
gas giant planets, we are conducting a spectroscopic search for 
planets orbiting a sample of metal-poor stars in the field, thus eliminating 
possible sources of interference with planet formation, or with migration 
to close-in orbits, or with planet survival, such as dynamical interactions 
in dense stellar environments. 

\section{Stellar Sample Selection}

In this project we are searching for planetary companions within 
1 AU orbiting a sample of 200 metal-poor dwarfs, chosen
from the Carney-Latham (e.g., Carney et al.\ 1994) and Ryan (Ryan
1989) surveys of metal-poor, chromospherically quiet, non rotating 
($V\sin i \leq 10$ km/s), high-velocity field 
stars that happen to be passing by the solar neighborhood. These stars 
are selected not to have close stellar companions based on long-term
radial-velocity (RV) monitoring that has been carried out with the CfA
Digital Speedometers. Our survey is orthogonal to other spectroscopic 
planet searches in that it spans a range of metallicities never 
investigated before ($-2.0\leq$ [Fe/H] $\leq -0.6$). We have further 
refined our sample of metal-poor dwarfs by imposing magnitude 
($V\leq 11.5$) and temperature ($T_{eff}\leq 6250$ K) cut-offs, and 
utilized the metallicity, temperature, and magnitude constraints to compute 
the relative exposure times needed to achieve 20 m/s RV 
precision for planet detection. This precision is sufficient to 
achieve sensitivity to RV variations due to planetary companions 
with minimum masses in the average 
range $0.59\leq M_{p}\leq 2.75$ M$_{J}$, or 
higher, for orbital periods in the range $0.01\leq P\leq 1$ yr. 
Our sample of 200 metal-poor stars should eventually provide a robust
3-$\sigma$ null result in the case of no detections.

\section{Results}

During the first two Keck 1/HIRES observing nights we have obtained spectra 
(1 template + 2 iodine exposures) for 40 stars covering a large range 
of metallicities and effective temperatures, with the primary aim of 
evaluating the RV precision as a function of effective temperature and
metallicity, and thus test the validity of our predictions for the 
dependence of the RV precision on these two parameters. The resulting 
RV difference distribution between the two nights 
exhibits an rms residual velocity $\sigma \simeq 25$ m/s, corresponding to an 
expected single-measurement uncertainty of 
$\varepsilon = \sigma/\sqrt{2} \simeq 18$ m/s. No clear trends 
in the RV differences as a function of [Fe/H] and T$_{eff}$
are present, a further confirmation that the model we developed for the 
dependence of the RV precision on the above parameters is robust. 
Our next goal will be to demonstrate the long-term stability of the velocity zero-point 
and expected single-measurement precision for planet detection.

\end{document}